\documentclass[applsci,article,accept,pdftex,moreauthors]{Definitions/mdpi} 
\firstpage{1} 
\pubvolume{12}
\issuenum{7}
\articlenumber{3293}
\pubyear{2022}
\copyrightyear{2022}
\externaleditor{{Academic Editor:} 
Sławomir K. Zieliński}
\datereceived{8 March 2022} 
\dateaccepted{22 March 2022} 
\datepublished{24 March 2022} 
\hreflink{https://doi.org/10.3390/\linebreak app12073293} 

\Title{You Only Hear Once: A YOLO-like Algorithm for Audio Segmentation and Sound Event Detection}

\TitleCitation{You Only Hear Once: A YOLO-like Algorithm for Audio Segmentation and Sound Event Detection}

\Author{{Satvik} 
 Venkatesh $^{1,}$*\orcidA{}, {David Moffat} $^{1,2}$\orcidB{} and Eduardo Reck Miranda $^{1}$\orcidC{}}

\AuthorNames{Satvik Venkatesh, David Moffat and Eduardo Reck Miranda}

\AuthorCitation{{Venkatesh,} 
 S.; Moffat, D.; Miranda, E.R.}

\address{%
$^{1}$ \quad Interdisciplinary Centre for Computer Music Research, University of Plymouth, Plymouth PL4 8AA, UK; dmof@pml.ac.uk (D.M.); eduardo.miranda@plymouth.ac.uk (E.R.M.)\\
$^{2}$ \quad Plymouth Marine Laboratory, Plymouth PL1 3DH, UK }

\corres{Correspondence: satvik.venkatesh@plymouth.ac.uk}

\abstract{Audio segmentation and sound event detection are crucial topics in machine listening that aim to detect acoustic classes and their respective boundaries. It is useful for audio-content analysis, speech recognition, audio-indexing, and music information retrieval. In recent years, most research articles adopt segmentation-by-classification. This technique divides audio into small frames and individually performs classification on these frames. In this paper, we present a novel approach called You Only Hear Once (YOHO), which is inspired by the YOLO algorithm popularly adopted in Computer Vision. We convert the detection of acoustic boundaries into a regression problem instead of frame-based classification. This is done by having separate output neurons to detect the presence of an audio class and predict its start and end points. The relative improvement for F-measure of YOHO, compared to the state-of-the-art Convolutional Recurrent Neural Network, ranged from 1\% to 6\% across multiple datasets for audio segmentation and sound event detection. As the output of YOHO is more end-to-end and has fewer neurons to predict, the speed of inference is at least 6 times faster than segmentation-by-classification. In addition, as this approach predicts acoustic boundaries directly, the post-processing and smoothing is about 7 times faster.}

\keyword{audio segmentation; sound event detection; you only look once; deep learning; regression; convolutional neural network; music-speech detection; convolutional recurrent neural network; radio} 

\begin{document}

%
%

\section{Introduction}
Audio segmentation and sound event detection have similar goals---to detect acoustic classes and their respective boundaries within an audio stream. They provide information regarding the content of audio and the temporal occurrences of audio events. It is helpful for indexing audio archives, target-based distribution of media, and as a pre-processing step for speech recognition \cite{butko2011audio}. In addition, detecting audio events in real-time is beneficial for self-driving automobiles \cite{dcase2021task4}, surveillance \cite{radhakrishnan2005audio}, bioacoustic monitoring \cite{salamon2016towards}, and intelligent remixing \cite{martinez2021deep}.

\textls[15]{The literature has commonly adopted two approaches to audio segmentation--- (1)~distance-based segmentation} and (2) segmentation-by-classification \cite{theodorou2014overview}. The first approach directly finds regions of high acoustic change through Euclidean distance or Bayesian information criterion \cite{huang2006advances}. The method divides audio into segments based on the peaks of acoustic change. Subsequently, the audio classes within each of these segments are detected. However, recent research has generally adopted the second approach, which is segmentation-by-classification. It presents sound event detection as a supervised learning task. This approach divides an audio file into frames, typically in the range of 10--25~ms, and classifies each frame individually. Effectively, we detect the onset and offset of each audio event by classification. 


Data to train a machine learning model for event detection require precise labels that mention the acoustic boundaries and classes. Annotating such datasets is a time-consuming and expensive process. Therefore, researchers have explored data-centric approaches such as artificial data synthesis to generate large-scale training data \cite{venkatesh2021artificially, salamon2017scaper}. Furthermore, researchers have explored weak label and semi-supervised learning to tackle the scarcity of labelled data \cite{turpault2019sound, miyazaki2020weakly}. Datasets such as AudioSet~\cite{gemmeke2017audio} are annotated with weak labels, which indicates that a sound event is present in the audio clip, but does not specify the timing within the audio. Hershey et al.~\cite{hershey2021benefit} emphasised the benefit of temporally strong labels to improve the performance of audio classifiers.

The architectures for audio segmentation have evolved from traditional machine learning models such as the Gaussian mixture model to deep neural networks. Bidirectional Long short-term memory (B-LSTM) networks have been effective in segmenting temporal data \cite{gimeno2020multiclass}. Lemaire et al. \cite{lemaire2019temporal} showed that the non-causal temporal convolutional neural network was more effective than the B-LSTM. However, the Convolutional Recurrent Neural Network (CRNN) obtains state-of-the-art performance on many sound event detection datasets because it combines the advantage of 2D convolutions and recurrent layers \cite{cakir2017convolutional, venkatesh2021investigating}.

There has been a growing interest in the community to adopt end-to-end deep learning for information retrieval from audio. Raw audio waveforms have been explored instead of features such as mel spectrograms for the input \cite{dieleman2014end, lee2017raw}. However, with regards to an end-to-end setup, there has been less attention given to the output of such networks. Traditionally, in segmentation-by-classification, the neural network classifies each audio frame. Subsequently, a post-processing step converts the neural network's output into human-readable labels. The disadvantage is that this post-processing is slow because each audio frame has to be serially processed. Therefore, in an ideal end-to-end setup, the neural network would output human-readable labels by directly predicting the boundaries of acoustic classes.

In order to output human-readable labels directly, sound event detection must be transformed from a classification problem to a regression problem. Phan~et~al.~\cite{phan2014random} proposed random regression forests for sound event detection and classification. Xu et al.~\cite{xu2014regression} adopted a regression approach for speech enhancement. However, most studies in the literature adopt frame-based classification, where the neural network classifies each frame separately. In this study, we present a novel neural network architecture inspired by the You Only Look Once (YOLO) algorithm \cite{redmon2016you}. YOLO gained attention in the Computer Vision community for object detection. It transformed bounding box prediction from a classification problem to a regression one. Using this approach, it obtained speedups of around 3$\times$ without compromising accuracy. 
We present a system called You Only Hear Once (YOHO) that predicts the boundaries of acoustic classes through regression.

YOLO has been adopted in the audio domain by visualising spectrograms as images. Zseb{\H{o}}k et al.~\cite{zsebHok2019automatic} adopted YOLO for automatic bird song and syllable segmentation. Segal~et~al.~\cite{segal2019speechyolo} presented a system called SpeechYOLO which treated audio fragments as objects. They adopted YOLO for keyword spotting tasks. Algabri et al. \cite{algabri2020towards} investigated object detection techniques such as YOLO and CenterNet \cite{zhou2019objects} for phoneme recognition. However, the novelty of the YOHO paradigm is that it converts frame-based classification into a regression problem by gradually reducing the temporal dimension through many convolutional layers. This makes the output of the network closer to human-readable labels, therefore reducing the need for post-processing. Separate neurons were used to detect the onset and offset of audio classes. We apply our system to audio segmentation and sound event detection tasks, where the literature has predominantly used frame-based classification. Furthermore, we present a multi-output system, which detects acoustic classes that can overlap with each other.

We evaluate the YOHO algorithm for multiple audio event detection tasks. First, we explore music-speech detection in broadcast signals. We also compare our results with state-of-the-art algorithms on the Music Information Retrieval Evaluation eXchange (MIREX) competition dataset 2018 \cite{mirex2018musicspeech}. Second, we test our model on the TUT sound event detection dataset, which represents common sounds related to human presence and traffic. It was the dataset used in the Detection and Classification of Acoustic Scenes and Events (DCASE) competition 2017 \cite{mesaros2017dcase}. Third, we evaluate our model on the Urban-SED dataset~\cite{salamon2017scaper}, which is a synthetic dataset for environmental audio. In all three cases, the YOHO algorithm performed better and faster than the CRNN. All the code associated with this project is available in this GitHub repository~(\href{https://github.com/satvik-venkatesh/you-only-hear-once}{https://github.com/satvik-venkatesh/you-only-hear-once}, accessed on 2 March 2022).

\section{You Only Hear Once (YOHO)}
\subsection{Motivation}
\begin{figure}[H]
	\includegraphics[width=.95\linewidth]{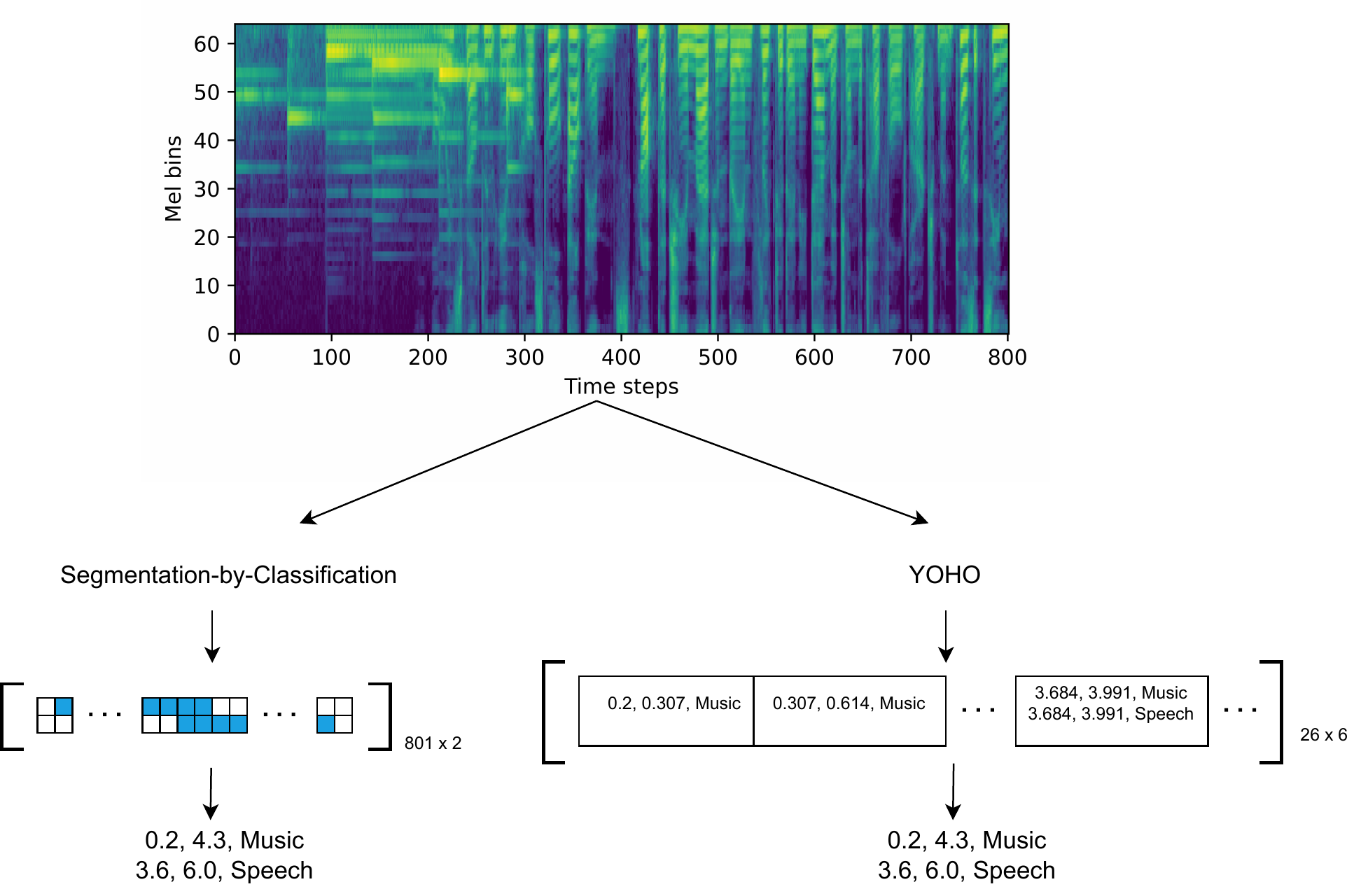}
	\caption{A comparison of segmentation-by-classification and YOHO. \label{fig:yoho-motivation}}
\end{figure}

In this paper, we intend to make the neural network output labels that are closer to human-readable labels. This way we make the pipeline more end-to-end. Figure \ref{fig:yoho-motivation} illustrates a comparison between segmentation-by-classification and the YOHO paradigm. For both paradigms, a mel spectrogram of shape $801\times 64$ is fed as input. In segmentation-by-classification, each time step is classified as music, speech, both, or none. Subsequently, these classifications are converted to human readable labels. However, in YOHO, each block of 0.307~s is processed through regression. One neuron detects the presence of an acoustic class. If the class is present, one neuron predicts the start point of the class and one neuron detects the end point of the class. Subsequently, during post-processing, these blocks of 0.307~s are merged to form a final prediction. Using this technique, the number of time steps is reduced from 801 to 26, which makes the network significantly faster, generalise better, and more end-to-end. More details on the implementation are given in the below subsections.

\vspace{-6pt}

\subsection{Network Architecture}
\label{sec:yoho-arch}
The network architectures used by different versions of YOLO \cite{redmon2016you, redmon2017yolo9000} were large and not suitable for our smaller training datasets. Therefore, we adapted the MobileNet architecture \cite{howard2017mobilenets} for our task. We modified the final layers of MobileNet to realize the YOHO algorithm. MobileNet has also been employed for audio classification by YamNet~\cite{yamnet}, which only detects audio classes, but not their segmentation boundaries.

As shown in Table \ref{table:arch}, YOHO is purely a convolutional neural network (CNN). We divide the table into two parts---the upper half comprising the original layers of the MobileNet architecture and the bottom half containing the layers that we have added. We use log-mel spectrograms as input features. The input dimension depends on the duration of the audio example and specifications of the mel spectrogram. Here, we explain the network for music-speech detection, whose input contains 801 times steps and 64 frequency bins. After reshaping the mel spectrogram to 801 $\times$ 64 $\times$ 1, we perform a 2D convolution with a stride of 2. Hence, the time dimension and frequency dimension are reduced by half. The MobileNet architecture uses many depthwise-separable convolutions \cite{sifre2014rigid} with \mbox{3 $\times$ 3 filters} followed by pointwise convolutions with 1 $\times$ 1 filters. All convolutions except the final layer were fitted with ReLu activations and batch normalization \cite{ioffe2015batch}. Each time we adopt a stride of 2, there is a reduction in the time and frequency dimensions. As shown in the lower half of Table \ref{table:arch}, we gradually reduce the number of filters from 1024 to 256.

\begin{table}[H] 
	\caption{The neural network architecture for YOHO. The upper half of the table comprises the original layers of MobileNet. The bottom half contains the layers that we have added. Conv2D and Conv1D stand for 2D and 1D convolutions, respectively. The convolutions use a stride of 1 unless mentioned otherwise and {`dw'} stands for depthwise convolution.\label{table:arch}} 
	\newcolumntype{C}{>{\centering\arraybackslash}X}
\centering
\begin{tabular*}{\hsize}{@{}@{\extracolsep{\fill}}clccc@{}}
	\toprule
	\multicolumn{2}{c}{\textbf{Layer Type}}                                                                 & \multicolumn{1}{c}{\textbf{Filters}}                                & \multicolumn{1}{c}{\textbf{Shape/Stride}}                               & \multicolumn{1}{c}{\textbf{Output Shape}}                                               \\ \midrule
	\multicolumn{2}{c}{Reshape}                                                                             & \multicolumn{1}{c}{-}                                               & \multicolumn{1}{c}{-}                                                     & \multicolumn{1}{c}{801 $\times$ 64 $\times$ 1}                                                        \\ \midrule
	\multicolumn{2}{c}{Conv2D}                                                                             & \multicolumn{1}{c}{32}                                              & \multicolumn{1}{c}{3 $\times$ 3/2}                                             & \multicolumn{1}{c}{401 $\times$ 32 $\times$ 32}                                                       \\ \midrule
	\multicolumn{2}{c}{Conv2D-dw}                                                                           & \multicolumn{1}{c}{-}                                               & \multicolumn{1}{c}{3 $\times$ 3}                                                 & \multicolumn{1}{c}{401 $\times$ 32 $\times$ 32}                                                       \\ \midrule
	\multicolumn{2}{c}{Conv2D}                                                                             & \multicolumn{1}{c}{64}                                              & \multicolumn{1}{c}{1 $\times$ 1}                                                 & \multicolumn{1}{c}{401 $\times$ 32 $\times$ 64}                                                       \\ \midrule
	\multicolumn{2}{c}{Conv2D-dw}                                                                           & \multicolumn{1}{c}{-}                                               & \multicolumn{1}{c}{3 $\times$ 3/2}                                             & \multicolumn{1}{c}{201 $\times$ 16 $\times$ 64}                                                       \\ \midrule
	\multicolumn{2}{c}{Conv2D}                                                                             & \multicolumn{1}{c}{128}                                             & \multicolumn{1}{c}{1 $\times$ 1}                                                 & \multicolumn{1}{c}{201 $\times$ 16 $\times$ 128}                                                      \\ \midrule
	\multicolumn{2}{c}{Conv2D-dw}                                                                           & \multicolumn{1}{c}{-}                                               & \multicolumn{1}{c}{3 $\times$ 3}                                                 & \multicolumn{1}{c}{201 $\times$ 16 $\times$ 128}                                                      \\ \midrule
	\multicolumn{2}{c}{Conv2D}                                                                             & \multicolumn{1}{c}{128}                                             & \multicolumn{1}{c}{1 $\times$ 1}                                                 & \multicolumn{1}{c}{201 $\times$ 16 $\times$ 128}                                                      \\ \midrule
	\multicolumn{2}{c}{Conv2D-dw}                                                                           & \multicolumn{1}{c}{-}                                               & \multicolumn{1}{c}{3 $\times$ 3/2}                                             & \multicolumn{1}{c}{101 $\times$ 8 $\times$ 128}                                                       \\ \midrule
	\multicolumn{2}{c}{Conv2D}                                                                             & \multicolumn{1}{c}{256}                                             & \multicolumn{1}{c}{1 $\times$ 1}                                                 & \multicolumn{1}{c}{101 $\times$ 8 $\times$ 256}                                                       \\ \midrule
	\multicolumn{2}{c}{Conv2D-dw}                                                                           & \multicolumn{1}{c}{-}                                               & \multicolumn{1}{c}{3 $\times$ 3}                                                 & \multicolumn{1}{c}{101 $\times$ 8 $\times$ 256}                                                       \\ \midrule
	\multicolumn{2}{c}{Conv2D}                                                                              & \multicolumn{1}{c}{256}                                             & \multicolumn{1}{c}{1 $\times$ 1}                                                 & \multicolumn{1}{c}{101 $\times$ 8 $\times$ 256}                                                       \\ \midrule
	\multicolumn{2}{c}{Conv2D-dw}                                                                           & \multicolumn{1}{c}{-}                                               & \multicolumn{1}{c}{3 $\times$ 3/2}                                             & \multicolumn{1}{c}{51 $\times$ 4 $\times$ 256}                                                        \\ \midrule
	\multicolumn{2}{c}{Conv2D}                                                                              & \multicolumn{1}{c}{512}                                             & \multicolumn{1}{c}{1 $\times$ 1}                                                 & \multicolumn{1}{c}{51 $\times$ 4 $\times$ 256}                                                        \\ \midrule
	\multicolumn{1}{c}{5$\times$} & \multicolumn{1}{l}{\begin{tabular}[c]{@{}c@{}}Conv2D-dw\\ Conv2D\end{tabular}} & \multicolumn{1}{c}{\begin{tabular}[c]{@{}c@{}}-\\ 512\end{tabular}} & \multicolumn{1}{c}{\begin{tabular}[c]{@{}c@{}}3 $\times$ 3\\ 1 $\times$ 1\end{tabular}} & \multicolumn{1}{c}{\begin{tabular}[c]{@{}c@{}}51 $\times$ 4 $\times$ 256\\ 51 $\times$ 4 $\times$ 256\end{tabular}} \\ \midrule
	\multicolumn{2}{c}{Conv2D-dw}                                                                           & \multicolumn{1}{c}{-}                                               & \multicolumn{1}{c}{3 $\times$ 3/2}                                             & \multicolumn{1}{c}{26 $\times$ 2 $\times$ 512}                                                        \\ \midrule
	\multicolumn{2}{c}{Conv2D}                                                                              & \multicolumn{1}{c}{1024}                                            & \multicolumn{1}{c}{1 $\times$ 1}                                                 & \multicolumn{1}{c}{26 $\times$ 2 $\times$ 1024}                                                       \\ \midrule
	\multicolumn{2}{c}{Conv2D-dw}                                                                           & \multicolumn{1}{c}{-}                                               & \multicolumn{1}{c}{3 $\times$ 3}                                                 & \multicolumn{1}{c}{26 $\times$ 2 $\times$ 1024}                                                       \\ \midrule
	\multicolumn{2}{c}{Conv2D}                                                                              & \multicolumn{1}{c}{1024}                                            & \multicolumn{1}{c}{1 $\times$ 1}                                                 & \multicolumn{1}{c}{26 $\times$ 2 $\times$ 1024}                                                       \\ \midrule
	\multicolumn{1}{l}{}    &                                                                                 & \multicolumn{1}{l}{}                                                 & \multicolumn{1}{l}{}                                                       & \multicolumn{1}{l}{}                                                                     \\
	\multicolumn{1}{l}{}    &                                                                                 & \multicolumn{1}{l}{}                                                 & \multicolumn{1}{l}{}                                                       & \multicolumn{1}{l}{}                                                                     \\ \midrule
	\multicolumn{2}{c}{Conv2D-dw}                                                                           & \multicolumn{1}{c}{-}                                               & \multicolumn{1}{c}{3 $\times$ 3}                                                 & \multicolumn{1}{c}{26 $\times$ 2 $\times$ 1024}                                                       \\ \midrule
	\multicolumn{2}{c}{Conv2D}                                                                              & \multicolumn{1}{c}{512}                                             & \multicolumn{1}{c}{1 $\times$ 1}                                                 & \multicolumn{1}{c}{26 $\times$ 2 $\times$ 512}                                                        \\ \midrule
	\multicolumn{2}{c}{Conv2D-dw}                                                                           & \multicolumn{1}{c}{-}                                               & \multicolumn{1}{c}{3 $\times$ 3}                                                 & \multicolumn{1}{c}{26 $\times$ 2 $\times$ 512}                                                        \\ \midrule
	\multicolumn{2}{c}{Conv2D}                                                                              & \multicolumn{1}{c}{256}                                             & \multicolumn{1}{c}{1 $\times$ 1}                                                 & \multicolumn{1}{c}{26 $\times$ 2 $\times$ 256}                                                        \\ \midrule
	\multicolumn{2}{c}{Conv2D-dw}                                                                           & \multicolumn{1}{c}{-}                                               & \multicolumn{1}{c}{3 $\times$ 3}                                                 & \multicolumn{1}{c}{26 $\times$ 2 $\times$ 256}                                                        \\ \midrule
	\multicolumn{2}{c}{Conv2D}                                                                              & \multicolumn{1}{c}{128}                                             & \multicolumn{1}{c}{1 $\times$ 1}                                                 & \multicolumn{1}{c}{26 $\times$ 2 $\times$ 128}                                                        \\ \midrule
	\multicolumn{2}{c}{Reshape}                                                                             & \multicolumn{1}{c}{-}                                               & \multicolumn{1}{c}{-}                                                     & \multicolumn{1}{c}{26 $\times$ 256}                                                            \\ \midrule
	\multicolumn{2}{c}{Conv1D}                                                                              & \multicolumn{1}{c}{6}                                               & \multicolumn{1}{c}{1}                                                     & \multicolumn{1}{c}{26 $\times$ 6}                                                              \\ 
	\bottomrule
\end{tabular*}
\end{table}

\begin{figure}[H]
	\includegraphics[width=0.65\linewidth]{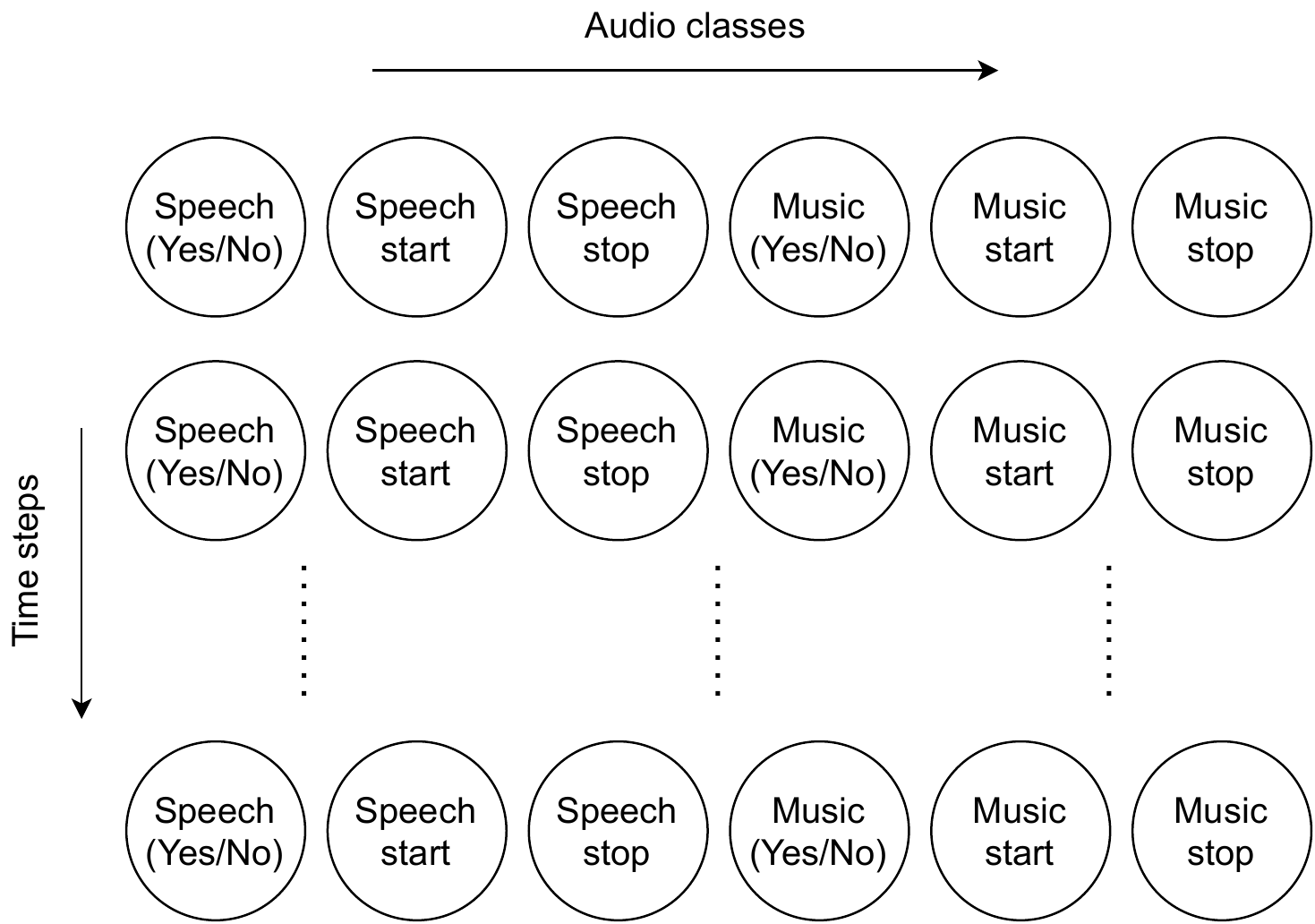}
	\caption{An illustration of the output layer of the YOHO algorithm. This network is for music-speech detection. To increase the number of audio classes, we add neurons along the horizontal axis.}
	\label{fig:output-layer}
\end{figure}

Subsequently, we flatten the last two dimensions. The final layer is a 1D convolution with six filters. The output shape is 26 $\times$ 6, where 26 stands for the number of time steps. This layer is similar to a convolutional implementation of sliding windows \cite{sermanet2014overfeat} along the time dimension. At each time step, the first neuron performs a binary classification that detects the presence of an acoustic class. The second and third neurons perform regression for the start and endpoints for the respective acoustic class. Figure \ref{fig:output-layer} illustrates the output layer of the YOHO algorithm.
\vspace{-4pt}

In this context, we are dealing with two acoustic classes---music and speech. Therefore, the output has six neurons at each time step. For example, if the length of an audio example is 8~s, each time step in the output corresponds to 0.307~s because there are 26 divisions. We applied sigmoid activations for all neurons in the output layer. Hence, we normalized the regression outputs between 0 and 1. Moreover, even if the input shape of the neural network is different, for example $257\times 40$, the neural network and the parameters of convolutional layers still remain exactly the same. The only difference would be the output shape of the neural network, which depends on the number of time steps in the input and the number of unique audio classes in the output.

\subsection{Loss Function}
Generally, neural networks such as the CRNN that use segmentation-by-classification adopt binary cross-entropy as the loss function. As we modeled the problem as a regression one, we used the sum squared error. Equation (1) shows the loss function for each acoustic class $c$.
\begin{equation}
	\mathcal{L\textsubscript{c}}(\hat{y}, y) = \begin{cases}
		(\hat{y\textsubscript{1}} - y\textsubscript{1})^2 + \\(\hat{y\textsubscript{2}} - y\textsubscript{2})^2 + 
		(\hat{y\textsubscript{3}} - y\textsubscript{3})^2,& \text{if } y\textsubscript{1} = 1\\
		(\hat{y\textsubscript{1}} - y\textsubscript{1})^2,              & \text{if } y\textsubscript{1} = 0
	\end{cases}
\end{equation}
where $y$ and $\hat{y}$ are the ground-truth and predictions respectively. $y\textsubscript{1} = 1$ if the acoustic class is present and $y\textsubscript{1} = 0$ if the class is absent. $y\textsubscript{2}$ and $y\textsubscript{3}$, which are the start and endpoints for each acoustic class are considered only if $y\textsubscript{1} = 1$. In other words, $(\hat{y\textsubscript{1}} - y\textsubscript{1})^2$ corresponds to the classification loss and $(\hat{y\textsubscript{2}} - y\textsubscript{2})^2 + (\hat{y\textsubscript{3}} - y\textsubscript{3})^2$ corresponds to the regression loss. The total loss $\mathcal{L}$ is summed across all acoustic classes.

\subsection{Example of Labels}
Table \ref{table:example-labels} shows an example of the output for the YOHO algorithm. The total length of the audio is 8 s. Within the example, Music occurs from 0.2 to 4.3 s and Speech occurs from 3.6 to 6.0 s. Note that each row in Table \ref{table:example-labels} corresponds to one time step, which is equal to 0.307~s. In addition, the regression values are normalized from 0 to 1. For example, if music starts at 0.2 s, the value is divided by 0.307 to get 0.65 as shown in the first row of Table \ref{table:example-labels}. 

\begin{table}[H] 
	\caption{An example of labels for the YOHO algorithm. Music occurs from 0.2 to 4.3 s and Speech occurs from 3.6 to 6.0 s. Note that start and stop values are considered only when the respective audio class is present. The dimensions of the output are 26 $\times$ 6. Note that each time step/row in the table corresponds to 0.307~s. The start and stop values are normalised on the range of 0 to 1. For instance, in the first time step, music's start point would be rescaled from 0.2 to 0.65. \label{table:example-labels}}
	\newcolumntype{C}{>{\centering\arraybackslash}X}
	\centering
	\begin{tabularx}{\textwidth}{CCCCCC}
	\toprule
	\textbf{\begin{tabular}[c]{@{}c@{}}Speech\\ (Yes/No)\end{tabular}} & \textbf{\begin{tabular}[c]{@{}c@{}}Speech\\ Start\end{tabular}} & \textbf{\begin{tabular}[c]{@{}c@{}}Speech\\ Stop\end{tabular}} & \textbf{\begin{tabular}[c]{@{}c@{}}Music\\ (Yes/No)\end{tabular}} & \textbf{\begin{tabular}[c]{@{}c@{}}Music\\ Start\end{tabular}} & \textbf{\begin{tabular}[c]{@{}c@{}}Music\\ Stop\end{tabular}} \\ \midrule
	0                                                                    & -                                                               & -                                                              & 1                                                                   & 0.65                                                           & 1.0                                                           \\
	0                                                                    & -                                                               & -                                                              & 1                                                                   & 0.0                                                            & 1.0                                                           \\
	0                                                                    & -                                                               & -                                                              & 1                                                                   & 0.0                                                            & 1.0                                                           \\
	0                                                                    & -                                                               & -                                                              & 1                                                                   & 0.0                                                            & 1.0                                                           \\
	0                                                                    & -                                                               & -                                                              & 1                                                                   & 0.0                                                            & 1.0                                                           \\
	0                                                                    & -                                                               & -                                                              & 1                                                                   & 0.0                                                            & 1.0                                                           \\
	0                                                                    & -                                                               & -                                                              & 1                                                                   & 0.0                                                            & 1.0                                                           \\
	0                                                                    & -                                                               & -                                                              & 1                                                                   & 0.0                                                            & 1.0                                                           \\
	0                                                                    & -                                                               & -                                                              & 1                                                                   & 0.0                                                            & 1.0                                                           \\
	0                                                                    & -                                                               & -                                                              & 1                                                                   & 0.0                                                            & 1.0                                                           \\
	0                                                                    & -                                                               & -                                                              & 1                                                                   & 0.0                                                            & 1.0                                                           \\
	1                                                                    & 0.7                                                             & 1.0                                                            & 1                                                                   & 0.0                                                            & 1.0                                                           \\
	1                                                                    & 0.0                                                             & 1.0                                                            & 1                                                                   & 0.0                                                            & 1.0                                                           \\
	1                                                                    & 0.0                                                             & 1.0                                                            & 1                                                                   & 0.0                                                            & 0.975                                                         \\
	1                                                                    & 0.0                                                             & 1.0                                                            & 0                                                                   & -                                                              & -                                                             \\
	1                                                                    & 0.0                                                             & 1.0                                                            & 0                                                                   & -                                                              & -                                                             \\
	1                                                                    & 0.0                                                             & 1.0                                                            & 0                                                                   & -                                                              & -                                                             \\
	1                                                                    & 0.0                                                             & 1.0                                                            & 0                                                                   & -                                                              & -                                                             \\
	1                                                                    & 0.0                                                             & 1.0                                                            & 0                                                                   & -                                                              & -                                                             \\
	1                                                                    & 0.0                                                             & 0.5                                                            & 0                                                                   & -                                                              & -                                                             \\
	0                                                                    & -                                                               & -                                                              & 0                                                                   & -                                                              & -                                                             \\
	0                                                                    & -                                                               & -                                                              & 0                                                                   & -                                                              & -                                                             \\
	0                                                                    & -                                                               & -                                                              & 0                                                                   & -                                                              & -                                                             \\
	0                                                                    & -                                                               & -                                                              & 0                                                                   & -                                                              & -                                                             \\
	0                                                                    & -                                                               & -                                                              & 0                                                                   & -                                                              & -                                                             \\
	0                                                                    & -                                                               & -                                                              & 0                                                                   & -                                                              & -                                                            \\
	\bottomrule
\end{tabularx}
\end{table}
\unskip

\subsection{Other Details}
We trained the network with the Adam optimizer, a learning rate of 0.001, a batch size of 32, and early stopping \cite{yao2007early}. In some cases, we used L2 normalization, spatial dropout, and SpecAugment \cite{park2017specaugment}. We used log-mel spectrograms as features for the neural network. The parameters of spectrograms were unique for each dataset. Section \ref{sec:datasets} contains the details for each case.

To evaluate the systems, we adopted the sed\_eval toolbox \cite{mesaros2016metrics}, which is common in the literature for audio segmentation and sound event detection. The python toolbox is openly available (\url{https://tut-arg.github.io/sed_eval/}, accessed on 17 March 2022) and presents a convenient interface to calculate metrics such as overall F-measure, error rate, class-based F-measures, and so on. The specifications of segment-based metrics for experiments in this paper are mentioned along with the relevant results in Section \ref{sec:results}.

\subsection{Post-Processing}
For music-speech detection, the output of the CRNN would be 801 $\times$ 2, corresponding to 801 times steps and two acoustic classes. On the other hand, the output for the YOHO network is 26 $\times$ 6. A post-processing step parses the output of the neural network to create human-readable labels. Subsequently, smoothing is performed over the output to eliminate the occurrence of spurious audio events. Two smoothing approaches are common in the literature---median filtering \cite{gimeno2020multiclass} and threshold-dependent smoothing \cite{lemaire2019temporal}. We adopted the latter approach. In this technique, if the duration of the audio event is too short or if the silence between consecutive events of the same acoustic class is too short, we remove the~occurrence. 

For music-speech detection, the minimum silence between consecutive music events or consecutive speech events was set to 0.8 s. The minimum duration for a music event was set to 3.4 s and for a speech event was 0.8 s. For environmental sound event detection, if the silence between consecutive audio events of the same acoustic class was less than 1.0~s, it was smoothed. We did not set any threshold for the minimum duration of an audio event for this task.

\subsection{Models for Comparison}
\label{sec:model-comp}
In this sub-section, we present two additional models, which are slight deviations from the YOHO architecture---CNN and CRNN. The motivation behind these models is to investigate which aspects of YOHO are actually advantageous. The feature-extraction for all these models is the same, which make them directly comparable. The CNN model aims to create a segmentation-by-classification version of the YOHO architecture. As you can see in Table \ref{table:arch}, some Conv2D-dw layers adopt a stride of [2, 2]. These strides were set to [1, 1] instead of [2, 2] and max-pooling was adopted to reduce the frequency dimension by half. This way, the time resolution of the network does not reduce through its depth. Note that using a stride of [1, 2] would have produced a similar effect of maintaining the time resolution and reducing the frequency resolution. However, TensorFlow currently does not support rectangular strides for depthwise convolutions and hence, we adopted max-pooling. The number of parameters in the CNN was 3.9 million, which is the same as the network for YOHO.

In the CRNN model, the first 13 layers were identical to the YOHO network. We skipped the convolutional layers where the number of filters became larger than 256~because the network became too large to fit into the RAM. Following the convolutional layers, we had two B-GRU layers with 80 units each. The number of parameters for the CRNN was 1.3~million, which is less than the YOHO network. Increasing the number of convolutional layers only worsened the performance of the CRNN. Therefore, it was optimal to have a CRNN with fewer parameters.

The output shape for the CNN and CRNN was $801 \times 2$, performing binary classification for music and speech at each time step. We compared the performance of YOHO with these two additional models on the in-house test set for music-speech detection. We also compared the inference times of these models. A summary of the architectures for comparison can be found in Table \ref{table:models-comparison}.

\begin{table}[H] 
	\caption{Models for comparison on the in-house test set for music-speech detection.\label{table:models-comparison}}
	\newcolumntype{C}{>{\centering\arraybackslash}X}
	\begin{tabularx}{\textwidth}{cC}
		\toprule
		\textbf{Model} & \textbf{Remarks} \\ 	
		\midrule
		YOHO                &    The architecture is explained in Section \ref{sec:yoho-arch}. \\ \midrule
		CNN                &    [2, 2] strides in convolutions are replaced by [1,~1]~strides, followed by max-pooling of [1,~2] to maintain the time resolution.   \\ \midrule
		CRNN               &      Only Conv2D and Conv2D-dw layers until 256~filters are included from Table \ref{table:arch}. After this, \textls[-15]{two B-GRU layers with 80 units each are added. }          \\
		\bottomrule
	\end{tabularx}
\end{table}

%
%
%
%
%
%
%
%


\section{Datasets}
\label{sec:datasets}
In this paper, we evaluate the robustness of the YOHO algorithm on multiple datasets. This section explains the different datasets and how we adapt the YOHO algorithm for each of them.

\subsection{Music-Speech Detection}
\label{sec:in-house-dataset}
Music-speech detection aims to detect the boundaries of music and speech in audio signals such as radio and TV programs. The neural network performs multi-output detection to allow the simultaneous occurrence of music and speech. The number of output neurons at each time step is six because we are detecting two acoustic classes. We obtained 5~h of audio from the MuSpeak dataset \cite{muspeak}. In addition, we collected 18~h of audio from BBC Radio Devon, which was manually annotated by the authors. Both datasets were roughly split into 50\% for training, 30\% for validation, and 20\% for testing.

There are many openly available datasets with separate files of music and speech, such as MUSAN \cite{snyder2015musan}, GTZAN \cite{tzanetakis2000marsyas, tzanetakis2002musical}, Scheirer and Slaney dataset \cite{scheirer1997construction}, and Instrument Recognition in Musical Audio Signals (IRMAS) \cite{bosch2012comparison}, to name a few. However, the problem with such datasets is that they are not mixed in the style of TV or radio programmes. Broadcast audio is generally well-mixed with instances of speech over background music, one song fading out and a new song fading in, and so on. In a previous study \cite{venkatesh2021investigating}, we presented an approach that artificially synthesises large training sets for music-speech detection. This technique automatically mixes separate files of music and speech in the style of a radio DJ. Various parameters such as audio fade curves and audio ducking are randomised to obtain a variety of synthetic examples. In the current paper, we included 46~h of synthetic examples in the training set. Table \ref{table:audio-seg-data} shows a brief overview of the contents of each split in the dataset. For a detailed explanation of the training sets and experimental setup, please refer to this study \cite{venkatesh2021investigating}.

\begin{table}[H] 
	\caption{Contents of train, validation, and test datasets for music-speech detection. Real-world radio data was collected from BBC Radio Devon and annotated by the authors. MuSpeak~\cite{muspeak} already contains annotations for music and speech. 46~h of artificial radio-like examples were synthesised by the method presented in this study \cite{venkatesh2021investigating}. \label{table:audio-seg-data}}
	\newcolumntype{C}{>{\centering\arraybackslash}X}
	\begin{tabularx}{\textwidth}{cC}
		\toprule
		\textbf{Dataset Division} & \textbf{Contents} \\ \midrule
		Train                &    46~h of synthetic radio data + 9~h from BBC Radio Devon + 1~h 30~{min} 
 from MuSpeak \\ \midrule
		Validation                &  5~h from BBC Radio Devon + 2~h from~MuSpeak \\  \midrule
		Test               &      4~h from BBC Radio Devon + 1~h 42~min from~MuSpeak       \\
		\bottomrule
	\end{tabularx}
\end{table}

All the audio files were resampled to 16~kHz. They were converted to mono by averaging the channels before pre-processing. Subsequently, we extracted 64 log-mel bins with a hop size of 10~ms and a window size of 25~ms. The frequencies for the mel spectrogram ranged from 125 Hz~to 7.5~kHz. We adopted audio features similar to those used by YamNet \cite{yamnet}. Note that we did not use any regularization such as L2 normalization, spatial dropout, or SpecAugment for this dataset because the training set is large.

We evaluate the model on two different test sets. The first one being our in-house test set, which contains approximately 4.5~h of audio from BBC Radio Devon and MuSpeak~\cite{muspeak}. The second one was the MIREX music-speech detection dataset, which contains 27~h of audio from various TV programs.

\subsection{TUT Sound Event Detection}
\label{sec:TUTSoundEventDetection}
The TUT Sound Event Detection dataset focuses on environmental sound detection~\cite{mesaros2017dcase}. It was adopted for the third task of the DCASE challenge 2017. It primarily consists of street recordings with traffic and other activity. Each audio example is 2.56~s. There were six unique audio classes---Brakes Squeaking, Car, Children, Large Vehicle, People Speaking, and People Walking. Thus, to predict the existence of the six classes, plus start and end times, we required 18 output neurons. The more recent DCASE challenges use additional techniques such as semi-supervised learning and source separation, which is not the focus of this study. Hence, we used the dataset from 2017 that contains only strongly labeled data. 

The total size of the dataset is approximately 1.5~h. The dataset comes with a four-fold cross-validation setup. The size of this dataset is significantly smaller than the one used for music-speech detection and may not be large enough for our deep learning architecture. Therefore, we applied L2 normalization of 0.001 on the first Conv2D layer. In addition, we included L2 normalization of 0.01 and spatial dropout of 0.1 on all the subsequent Conv2D layers. For data augmentation, we incorporated SpecAugment \cite{park2017specaugment}, which randomly drops a sequence of frequency bins or time steps from the input. Note that there were slight differences in our implementation of SpecAugment. We did not use any time warping because it becomes complicated to redefine labels for audio events. In addition, we applied SpecAugment on batches instead of individual examples to save computational time. 

The database contained stereo audio files with a sampling rate of 44.1 kHz. These were downmixed to mono before pre-processing. Subsequently, we extracted 40~log-mel bands in the range of 0 to 22,050 Hz. The hop size was 10~ms and the window size was 40~ms. We adopted audio features similar to the baseline system \cite{mesaros2017dcase} for the task, except that we used a smaller hop size. As the input of the network contains 2.56~s of audio, the input shape is 257 $\times$ 40 corresponding to 257 times steps and 40 mel bins. The output shape of the network is 9 $\times$ 18, corresponding to 9 times steps and 6 acoustic classes. Note that each time step in this case is 0.284~s, which is different from 0.307~s for music-speech detection. In both cases, we used the same network and the same sequence of convolutional layers. The convolutions with a stride of 2 reduces the temporal dimension by half. Hence, due to different input sizes, the number of time steps is 9 in one case and 26 in the other case. There were no special measures taken to estimate the duration of each time step beforehand. However, in most cases it was somewhere around 0.3~s, due to the hop and window sizes selected for feature extraction.

\subsection{Urban-SED}
The Urban Sound Event Detection dataset is a purely synthetic dataset generated by using scaper \cite{salamon2017scaper}. Each audio example was 10~s. There were ten unique audio classes---Air Conditioner, Car Horn, Children Playing, Dog Bark, Drilling, Engine Idling, Gun Shot, Jackhammer, Siren, and Street Music. The total size of the dataset is about 30~h and contains pre-defined splits for training, validation, and testing. As there were ten audio classes, the number of output neurons in YOHO was 30.

We used the same audio features as explained in Section \ref{sec:TUTSoundEventDetection}. For this dataset, we did not use any SpecAugment because the training set was larger. The L2 normalization and spatial dropout were identical to those used in Section \ref{sec:TUTSoundEventDetection}. As the input of the network contains 10 s of audio, the input shape is 1001 $\times$ 40 corresponding to 1001 times steps and 40 mel bins. The output shape of the network is 32 $\times$ 30, corresponding to 32 times steps and ten acoustic classes.

\section{Results}
\label{sec:results}

\subsection{Music-Speech Detection}
\subsubsection{In-House Test Set}
Table \ref{table:in-house-music-speech} shows the results on our in-house test set. F-measure was calculated using the sed\_eval \cite{mesaros2016metrics} module with a segment size of 10~ms. We compare the results of YOHO with the CNN and CRNN models explained in Section \ref{sec:model-comp}. In addition, we compare the performance with CNN and CRNN architectures  published in previous research~\cite{venkatesh2021artificially, venkatesh2021investigating}. 
All the deep learning models were trained using the same training set. YOHO obtains the highest F-measure for overall, music, and speech. YOHO significantly outperforms the CNN, which is the segmentation-by-classification version of the model. It is important to note that both models follow the same process for feature extraction and have the same number of parameters. This shows that our regression approach of predicting the acoustic boundaries directly is effective. The other CNN~\cite{venkatesh2021investigating} used larger kernel sizes such as 9 and 11, which may have improved the F-measure of Speech.

YOHO also outperforms the three CRNN architectures. CRNN~\cite{venkatesh2021artificially} used a kernel size of 7 and CRNN~\cite{venkatesh2021investigating} used kernel sizes
of 3, 11, and 11. In addition, CRNN~\cite{venkatesh2021investigating} used layer normalisation \cite{ba2016layer} instead of batch normalisation \cite{ioffe2015batch}. Therefore, we show that YOHO outperforms a variety of CRNN architectures in the literature. 

\begin{table}[H] 
	\caption{Results on our in-house test set for music-speech detection. The F-measures for overall, music, and speech are presented as percentages. {The values in bold indicate the largest number in each column.} \label{table:in-house-music-speech}}
	\newcolumntype{C}{>{\centering\arraybackslash}X}
	\begin{tabularx}{\textwidth}{CCCC}
		\toprule
		\textbf{Algorithm} & \textbf{F\textsubscript{overall}} & \textbf{F\textsubscript{music}} & \textbf{F\textsubscript{speech}} \\
		\midrule
YOHO               & \textbf{97.22}    & \textbf{98.20}   & \textbf{94.89} \\
CRNN               & {96.79}    & {97.84}   & {94.26} \\ 
CNN               & {93.89}    & {97.96}   & {85.13} \\ \midrule
CRNN \cite{venkatesh2021investigating}      & 96.37             & 97.37           & 94.00       \\
CRNN \cite{venkatesh2021artificially}      & 96.24             & 97.30            & 93.80 \\             
CNN \cite{venkatesh2021investigating}              & {95.23}    & {97.72}   & {89.62} \\
		\bottomrule
	\end{tabularx}
\end{table}
\unskip

\subsubsection{MIREX Music-Speech Detection}
Table \ref{table:mirex-music-speech} shows the results on the MIREX music-speech detection dataset. YOHO obtains the highest overall F-measure, which makes it the state-of-the-art for music-speech detection. The music F-measure for a CRNN \cite{venkatesh2021artificially} slightly surpassed the YOHO algorithm by 0.1\%. However, YOHO obtained the highest F-measure for speech. 

\begin{table}[H] 
	\caption{{Evaluation} 
 on the MIREX music-speech detection dataset 2018. The results of other studies were obtained from the MIREX website \cite{mirex2018musicspeech}. The F-measures are presented as percentages. {The values in bold indicate the largest number in each column.} \label{table:mirex-music-speech}}
	\newcolumntype{C}{>{\centering\arraybackslash}X}
	\begin{tabularx}{\textwidth}{CCCC}
		\toprule
	\textbf{Algorithm} & \textbf{F\textsubscript{overall}} & \textbf{F\textsubscript{music}} & \textbf{F\textsubscript{speech}} \\
		\midrule
YOHO               & \textbf{90.20}     & 85.66  & \textbf{93.18}   \\
CRNN \cite{venkatesh2021artificially}             & 89.53             & \textbf{85.76}           & 92.21            \\
CRNN \cite{venkatesh2021investigating}  & 89.09             & 85.01           & 92.16            \\
CNN \cite{marolt2018music}          & -                 & 54.78           & 90.9             \\
\textls[-15]{\mbox{Logistic Regression \cite{marolt2018music}}}             & -                 & 38.99           & 91.15            \\
ResNet \cite{marolt2018music}             & -                 & 31.24           & 90.86         \\  
MLP \cite{choi2018hybrid}              & -                 & 49.36           & 77.18            \\
		\bottomrule
	\end{tabularx}
\end{table}
\unskip

\subsection{TUT Sound Event Detection}
Table \ref{table:TUT-results} shows the results on the TUT sound event detection dataset. It also contains the results of the top three performers in the competition. For this competition, they adopted error rate \cite{mesaros2016metrics} as the main metric. Note that a lower error rate indicates better performance of the algorithm. Furthermore, a segment size of 1~s was adopted to calculate segment-based metrics. The first place in the competition was the CRNN architecture~\cite{adavanne2017report}. They~used $3\times3$ kernels followed by B-GRU layers with 32 units. Their model was optimised by a random hyper-parameter search~\cite{bergstra2012random} for number of layers and units. The second place in the competition adopted a multi-input CNN with $3\times 3$ kernels and a bespoke feature extraction process. The third place adopted a B-GRU model. Note that all three models adopt segmentation-by-classification. YOHO obtained a better error rate than the CRNN~\cite{adavanne2017report}, CNN~\cite{jeong2017detection}, and B-GRU~\cite{lu2017bidirectional} models. To ensure that the improvement in performance was not attributed to data augmentation, we re-trained the best CRNN network \cite{adavanne2017report} with SpecAugment. However, it worsened the performance of the algorithm. This may be because the CRNN uses segmentation-by-classification. Therefore, masking series of time steps leads to noise in the labels. However, the YOHO algorithm is relatively robust to this issue as it directly predicts boundaries through regression. 

Our results are not state-of-the-art on this dataset. Vesperini et al. \cite{vesperini2019polyphonic} adopted a Capsule Neural Network (CapsNet) and binaural short-time Fourier transform (STFT) for feature extraction and obtained an error rate 0.58. Luo et al. \cite{luo2021system} presented a Capsule Neural Network Recurrent Neural Network (CapsNet-RNN) that obtained an error rate of 0.57. However, these optimisations were beyond the scope of this study. It is important to note that YOHO is a paradigm and not an architecture. We show that regression outperforms segmentation-by-classification for multiple models. Future research can explore how YOHO can be optimised by adopting a CapsNet-style architecture.

\begin{table}[H] 
	\caption{Results on the TUT sound event detection dataset. {The value in bold indicates the algorithm with the lowest error rate.} \label{table:TUT-results}}
	\newcolumntype{C}{>{\centering\arraybackslash}X}
	\begin{tabularx}{\textwidth}{CC}
		\toprule
	\textbf{Algorithm} & \textbf{Error Rate} \\ 	
	\midrule
CapsNet-RNN \cite{luo2021system}               & \textbf{0.57}        \\
CapsNet \cite{vesperini2019polyphonic}               & {0.59}       \\
YOHO               & {0.75}                \\
CRNN \cite{adavanne2017report}              & 0.79        \\
CNN   \cite{jeong2017detection}             & 0.81         \\
B-GRU  \cite{lu2017bidirectional}            & 0.83                       \\
		\bottomrule
	\end{tabularx}
\end{table}
\unskip

\subsection{Urban-SED}

\begin{figure}[H]
	\includegraphics[width=0.9\linewidth]{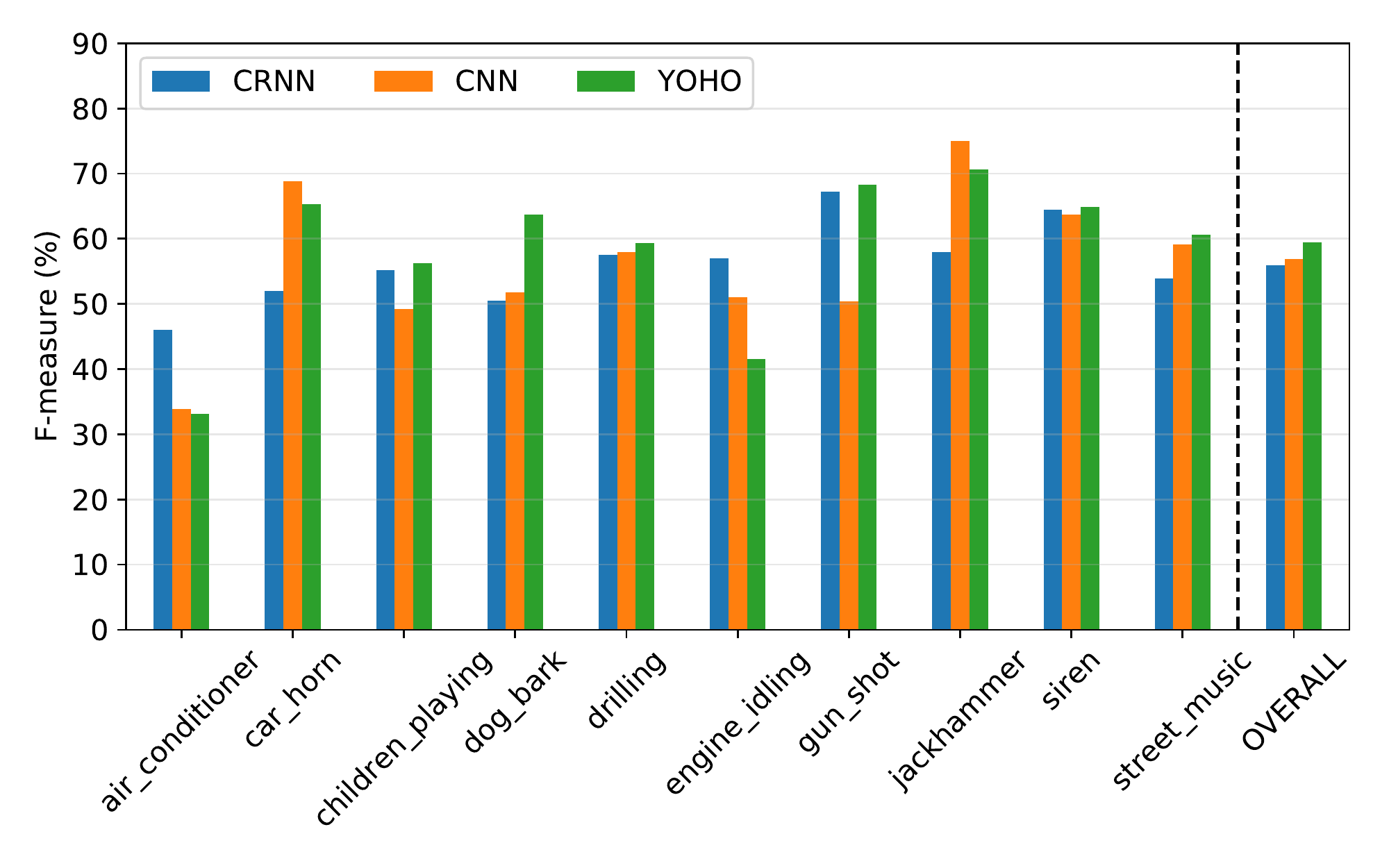}
	\caption{Segment-based F-measures for each class on the Urban-SED dataset calculated using segment-size of 1 s.\label{fig:urban-SED-results}}
\end{figure}

Table \ref{table:urban-SED-results} shows the results on the Urban-SED dataset for overall F-measure. A comparison of class-wise performance is also presented in Figure \ref{fig:urban-SED-results}. The YOHO algorithm is compared with the CRNN and CNN model presented by Salamon et al. \cite{salamon2017scaper}. YOHO obtains the highest overall F-measure. Among class-wise F-measures, YOHO obtains the highest for Children Playing, Dog Bark, Drilling, Gun Shot, Siren, and Street Music. CRNN obtains the highest for Air Conditioner and Engine Idling. CNN obtains the highest for Car Horn and Jackhammer. 

As you can see in Table \ref{table:urban-SED-results}, Mart{\'\i}n-Morat{\'o} et al. \cite{martin2019sound} adopted sound event envelope estimation on a CRNN model to improve the overall F-measure to 64.7\%, compared to 59.5\% obtained by YOHO. In future research, YOHO's performance can be improved by incorporating techniques like envelope estimation. In addition, weakly supervised sound event detection with envelope estimation has further improved the performance of the CRNN on this dataset \cite{dinkel2021towards}.

\begin{table}[H] 
	\caption{Segment-based overall F-measure on the Urban-SED dataset. {The value in bold indicates the algorithm with the highest F-measure.} \label{table:urban-SED-results}}
	\newcolumntype{C}{>{\centering\arraybackslash}X}
	\begin{tabularx}{\textwidth}{CC}
		\toprule
		\textbf{Algorithm} & \textbf{F\textsubscript{overall}} \\	\midrule
		CRNN with envelope estimation \cite{martin2019sound}               & \textbf{64.70}        \\
		YOHO & 59.50 \\
		CNN \cite{salamon2017scaper}               & {56.88}       \\
		CRNN \cite{salamon2017scaper}              & 55.96        \\
		\bottomrule
	\end{tabularx}
\end{table}

\subsection{Speed of Prediction}
\begin{figure}[H]
	\includegraphics[width=.98\linewidth]{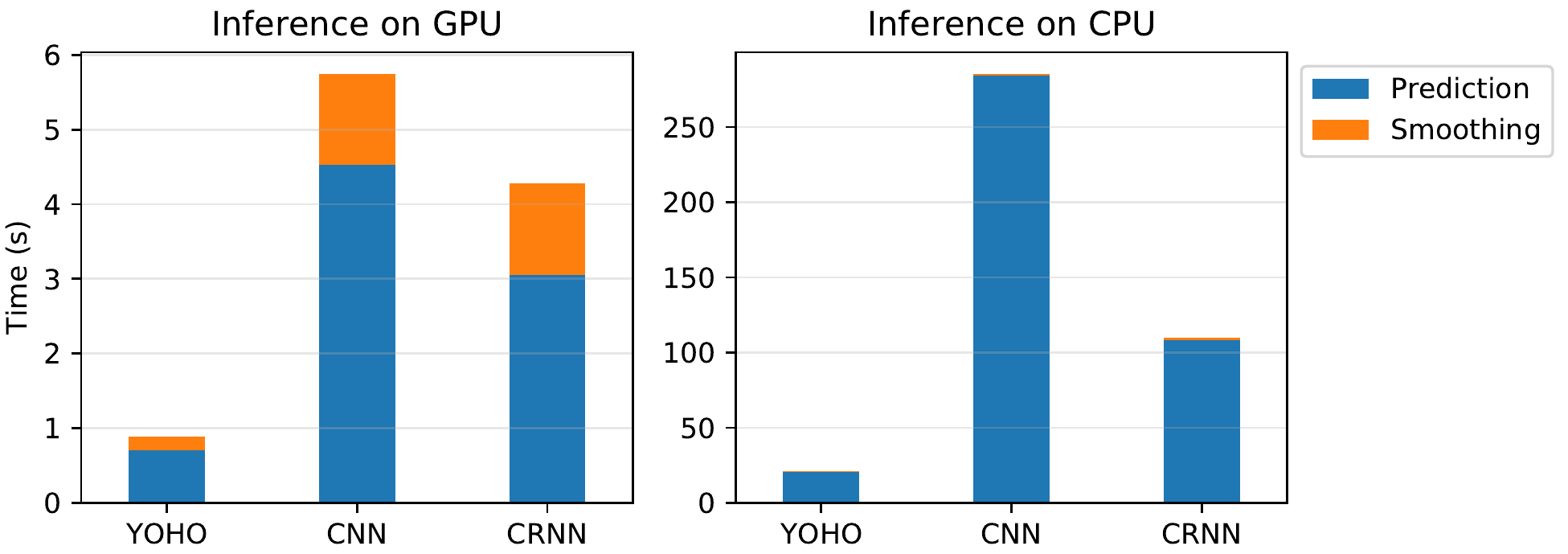}
	\caption{\textls[-15]{Average time taken to make predictions on 1~h of audio for music-speech detection. `Prediction' refers to the time taken by the network to make predictions. `Smoothing' is the post-processing step to parse the output of the network. The GPU used for inference was the Tesla P100.} \label{fig:time-results}}
\end{figure}

In this section, we compare the inference times of YOHO, CNN and CRNN models for music-speech detection. This experiment was performed on the in-house test set explained in Section \ref{sec:in-house-dataset}. To calculate the inference time, the prediction was made over the entire test set. Later, the inference time was divided by the number of hours of audio to obtain the average time taken per hour of audio. As we are adopting Google Colab for experiments, we ensured that all models were tested within the same runtime session. This way, we ensure that the same computing resources were given to YOHO, CNN, and CRNN. While training the models in earlier runtime sessions, we had stored their weights on Google Drive. When running the experiment to calculate inference times, these weights were loaded from Google Drive. Note that separate runtime sessions were used to calculate inference times over CPU and GPU as shown in Figure \ref{fig:time-results}, however the same session was used for inter-model comparison. Some important aspects of the system configuration were---Intel Xeon CPU processor, 12~GB~RAM, and Tesla P100 GPU (only when GPU was~used).


Figure \ref{fig:time-results} compares the inference times of YOHO, CNN and CRNN models for music-speech detection on the in-house test set. The CNN and CRNN models were explained in Section \ref{sec:model-comp}. YOHO and the CNN had exactly the same number of parameters, which is 3.9~million. The only difference is that the CNN adopts frame-based classification instead of regression. The CRNN model had 1.3 million parameters, which was less than the CNN and YOHO. On the CPU, the prediction time of YOHO was 14 times faster than the CNN and 5 times faster than the CRNN. On the Graphical Processing Unit (GPU), the prediction time of YOHO was 6 times faster than the CNN and 4 times faster than the CRNN. The increase in prediction speed is because YOHO has to predict only $26 \times 6$ neurons, whereas the CNN and CRNN have to predict $801 \times 2$ neurons. Despite the CRNN having fewer parameters, YOHO is significantly faster.

\vspace{-6pt}

As YOHO outputs outputs acoustic boundaries directly, the post-processing and smoothing for YOHO was 7 times faster than the CNN and CRNN. Note that the smoothing is performed only on the CPU.

\section{Discussion}
The results in Section \ref{sec:results} show that YOHO has multiple advantages over the state-of-the-art CRNN architecture. We examined the model for two different tasks---music-speech detection and environmental sound event detection. Music-speech detection is relatively a simpler task because of a larger and diverse training set. Additionally, there are only two acoustic classes to predict. On the other hand, environmental sound event detection was harder because of smaller and lower quality training sets. In addition, the number of acoustic classes was greater. However, in both scenarios, YOHO generalised better than CRNN and CNN. YOHO obtained state-of-the-art performance for music-speech detection on the MIREX 2018 competition dataset. We understand that YOHO has not obtained state-of-the-art performance on TUT Sound Events and Urban-SED datasets. However, it is important to note that the purpose of this study is to shift the paradigm from frame-based classification to regression for audio segmentation and sound event detection. There is a vast body of research involving CNN and CRNN architectures. It is not within the capacity of this study to incorporate all these optimisations for YOHO. As this is the first study that explores this paradigm, we believe that optimisations such as weak label learning \cite{miyazaki2020weakly} and envelope estimation \cite{martin2019sound} will improve YOHO's performance.

We also explored the idea of creating a regression-based CRNN that adopts the YOHO paradigm. We replaced the Conv1D layer with a B-GRU block. However, this slightly worsened the performance of the algorithm. This is because the YOHO network has many convolutional layers that reduces the temporal resolution from 801 to 26. Hence, the B-GRU blocks may not be effective on such a small number of time steps. However, alternative structures such as CNN-transformers \cite{kong2020sound} may be a promising avenue to explore.

YOHO was significantly quicker than the CNN and CRNN models because it had to predict fewer outputs and computationally cheaper post-processing. As explained in the paper, the output produced by YOHO is more end-to-end. For example, the output dimensions in music-speech detection is 26 $\times$ 6 for YOHO versus 801 $\times$ 2 for the CRNN. This corresponds to 156 output neurons for YOHO and 1602 for the CRNN. Furthermore, the CRNN needs to convert frame-based classifications to time boundaries. However, YOHO directly outputs the time boundaries. Due to the above reasons, YOHO is significantly quicker. Due to faster inference, YOHO is more suitable for real-time applications such as surveillance, self-driving automobiles, bioacoustic monitoring, and real-time remixing.



\section{Conclusions}
In this paper, we proposed a novel paradigm called YOHO for audio segmentation and sound event detection. It obtained state-of-the-art performance for music-speech detection and surpassed the CRNN and CNN's performance for environmental audio. YOHO presents sound event detection differently from the traditional segmentation-by-classification approach. We primarily adapted the MobileNet architecture \cite{howard2017mobilenets} to develop the YOHO paradigm. Future developments in the network architecture for YOHO would lead to improvements in performance. For instance, adding skip connections through ResNets \cite{he2016deep} or by including Inception blocks \cite{szegedy2015going}. Furthermore, there is scope to create hybrid architectures such as CNN-transformers \cite{kong2020sound} by adopting the YOHO paradigm.

Although YOHO's output is more end-to-end by predicting acoustic boundaries directly, it is limited by the time-resolution of the input, which is the mel spectrogram. It would be interesting to explore YOHO with raw audio, which would make the sound event detection pipeline completely end-to-end. Moreover, the YOHO approach is relevant to related tasks such as singing voice detection. Furthermore, recent studies have successfully combined sound event detection with source separation and semi-supervised learning~\mbox{\cite{turpault2019sound, turpault2021sound}.} Future work could explore how YOHO would perform in these scenarios.

\vspace{6pt} 


\clearpage
\authorcontributions{Conceptualization, S.V. and D.M.; methodology, S.V. and D.M.; software, S.V.; investigation, S.V.; writing---original draft preparation, S.V.; writing---review and editing, D.M. and S.V.; supervision, D.M. and E.R.M.; project administration, E.R.M.; funding acquisition, E.R.M. All authors have read and agreed to the published version of the manuscript.}

\funding{This study was supported by Engineering and Physical Sciences Research Council (EPSRC) grant EP/S026991/1.}

\institutionalreview{{Not applicable.} 
}

\informedconsent{{Not applicable.} 

}

\dataavailability{The data provided by BBC Radio Devon is copyrighted material and cannot be shared. Synthetic radio data for music-speech detection can be generated by using techniques in this paper \cite{venkatesh2021investigating}. MuSpeak \cite{muspeak} is an openly available annotated dataset for music-speech detection, which can be utilised by researchers for validation and testing. TUT Sound Events 2017 and Urban-SED datasets are publicly available. The code associated with this paper is openly available in this GitHub repository (\url{https://github.com/satvik-venkatesh/you-only-hear-once/}, accessed on 2~March 2022).} 

\acknowledgments{The authors would like to thank Blai Meléndez Catalán for helping us evaluate our model on the MIREX music-speech competition dataset. We thank Justin Salamon for providing insights on the Urban-SED dataset \cite{salamon2017scaper} and details of the results presented in Figure \ref{fig:urban-SED-results}. Research in this paper was conducted on Google Colab and we are thankful for their service.}

\conflictsofinterest{The authors declare no conflict of interest.}



\abbreviations{Abbreviations}{
The following abbreviations are used in this manuscript:\\

\noindent 
\begin{tabular}{@{}ll}
B-GRU & Bidirectional Gated Recurrent Unit\\
B-LSTM & Bidirectional Long Short-Term Memory \\BBC & British Broadcasting Corporation \\
CapsNet & Capsule Neural Network \\
CapsNet-RNN & Capsule Neural Network Recurrent Neural Network \\ 
CNN & Convolutional Neural Network\\
CRNN & Convolutional Recurrent Neural Network \\
DCASE & Detection and Classification of Acoustic Scenes and Events \\
GPU & Graphical Processing Unit \\
MIREX & Music Information Retrieval Evaluation eXchange \\
MLP & Multi-Layer Perceptron \\
RAM & Random Access Memory \\
STFT & Short-Time Fourier Transform\\
Urban-SED & Urban Sound Event Detection \\
YOHO & You Only Hear Once \\
YOLO & You Only Look Once \\
\end{tabular}}

%
%
%
\begin{adjustwidth}{-\extralength}{0cm}

\reftitle{References}

\end{adjustwidth}
\end{document}